# Conductive Domain Walls in Non-Oxide Ferroelectrics $Sn_2P_2S_6$

Jianming Deng[1,#], Xingan Jiang[1,#], Yanyu Liu[1], Wei Zhao[1], Yun Li[2], Ziyan Gao[1], Peng Lv[1], Sheng Xu[3], Tian-Long Xia[3], Jinchen Wang[3], Meixia Wu[4], Zishuo Yao[2], Xueyun Wang[1*], Jiawang Hong[1*]

[1]School of Aerospace Engineering, Beijing Institute of Technology, Beijing, 100081, China

[2]School of Chemistry and Chemical Engineering, Beijing Institute of Technology, Beijing, 100081, China

[3]Department of Physics and Beijing Key Laboratory of Opto-electronic Functional Materials & Micro-nano Devices, Renmin University of China, Beijing 100871, China

[4]Key Laboratory of Bioinorganic and Synthetic Chemistry of Ministry of Education, School of Chemistry, Sun Yat-Sen University, Guangzhou 510275, P. R. China.

[#]These authors contribute equally to this work.

[*]Corresponding e-mails: xueyun@bit.edu.cn, hongjw@bit.edu.cn

# ABSTRACT

The conductive domain wall (CDW) is extensively investigated in ferroelectrics, which can be considered as a quasi-two-dimensional reconfigurable conducting channel embedded into an insulating material. Therefore, it is highly important for the application of ferroelectric nanoelectronics. Hitherto, most CDW investigations are restricted in oxides, and limited work has been reported in non-oxides to the contrary. Here, by successfully synthesizing the non-oxide ferroelectric $Sn_2P_2S_6$ single crystal, we observed and confirmed the domain wall conductivity by using different scanning probe techniques which origins from the nature of inclined domain walls. Moreover, the domains separated by CDW also exhibit distinguishable electrical conductivity due to the interfacial polarization charge with opposite signs. The result provides a novel platform for understanding electrical conductivity behavior of the domains and domain walls in non-oxide ferroelectrics.

# 1. Introduction

The observation of nanometer sized electrically conductive domain wall (CDW) in insulating ferroelectrics has motivated the idea of designing domain-wall-based nanoelectronics [1], such as memeristive [2] and half-wave rectification devices [3]. CDW generally results from the substantial amount of bound charges due to the nature of head-to-head or tail-to-tail components of spontaneous polarization, and concentration of free carriers at the domain wall [4]. Hitherto, the reported CDW in ferroelectrics are mainly restrained in oxides, such as $BiFeO_3$, $BaTiO_3$, $Pb(Zr,Ti)O_3$, $LiNbO_3$, hexagonal manganites, $(Ca,Sr)_3Ti_2O_7$, *etc.*[5-11]. In these ferroelectric oxides, the oxygen defects play imperative roles in modulating the functionalities of domains and domain walls, such as the domain wall conductivity [12], especially when investigated by using scanning tip technique [13-15]. One circumventing strategy is finding CDW in non-oxide ferroelectrics. However, to the best of our knowledge, there is no report yet. In principle, there should be no such limitations in non-oxide ferroelectrics if there exists head-to-head or tail-to-tail components of spontaneous polarization. Therefore, exploring the CDW in non-oxide ferroelectrics are both fundamentally and practically rewarding, which may not only shed light on the intrinsic origin of the domain wall conductivity, but also pave the path for more stable domain-wall-based nanoelectronics, as well as tune the functionalities of domain wall using possible new mechanism rather than oxygen defects which is a common method in oxide ferroelectrics.

So far, only few inorganic non-oxide ferroelectrics has been experimentally realized, such as thiourea[16-17], $Rb_2ZnCl_4$[18], and metal seleno- and thio-phosphates[19-21], but no domain wall conductivity has been reported yet. Among all the inorganic non-oxide candidates, $CuInP_2S_6$ is one of the most widely investigated system [22-24], yet no CDW has been observed, possibly due to the van der Waals gap, preventing charge carriers from moving along the domain walls. Here, we choose a special metal-selenophosphate, $Sn_2P_2S_6$ (SPS), which has sulfur-phosphorus bonding in the van der Waals gap, to explore the possible CDW in non-oxide ferroelectric. At room temperature, SPS crystallizes in the space group *Pn* with a canted polarization [25,26], and a ferroelectric phase transition occurs near the $T_c$ = 339 K [27]. The fascinating physical phenomena in SPS such as the three-well local potential, and negative thermal expansion and so forth [28-30] can enable the realization of novel multifunctional devices.

In this work, we observed the nanoscale conduction at domain walls by a combination study of piezoresponse force microscopy (PFM), conductive atomic force microscopy (c-AFM) and scanning microwave impedance microscope (sMIM) in SPS single crystal, due to the nature of inclined domain walls. Furthermore, we revealed the coexistence of in-plane and out-of-plane ferroelectric polarizations. Interestingly, the two different domains with up and down polarization components perpendicular to the (011) surface exhibit distinguishable electrical conductivity from c-AFM measurement, which is caused by the different band bending behavior due to interfacial charge with opposite signs.

## 2. Results and discussion

The grown SPS crystal looks like a truncated prismoid, as shown in the inset of Figure 1a. As-grown SPS sample shows good stability against moisture and oxygen oxidation at ambient condition. Based on X-ray diffraction measurement, the top facet of the specimen is confirmed as (011) surface, as shown in Figure S1a. To confirm the homogeneity and the composition of SPS, energy dispersive spectrometer (EDS) analysis was carried out, as shown in Figure S1b in Supporting Information. The structural examinations confirmed the monoclinic *Pn* symmetry ($T<T_c$) and $P2_1/n$ symmetry ($T>T_c$) in SPS single crystal. The detailed atomic structure information can be found in Table 1 and Table S1.

The ferroelectric crystal structure of SPS is presented in Figure S1c. In conventional unit cell, SPS contains two $(P_2S_6)^{4-}$ anionic units, whereas four $Sn^{2+}$ cations are arranged between these anionic units. Each $(P_2S_6)^{4-}$ unit consists of two distorted trigonal $PS_3$ pyramids bound by a P-P bond.

Table 1 Temperature-dependence lattice parameters of SPS from X-ray diffraction measurement.

| Temperature (K) | Space group | Lattice parameter | | | |
|---|---|---|---|---|---|
| | | $a$ (Å) | $b$ (Å) | $c$ (Å) | $\beta$ (°) |
| 163 | $Pn$ | 6.5125(13) | 7.4760(15) | 9.3801(19) | 91.15(3) |
| 300 | $Pn$ | 6.5297(13) | 7.4839(15) | 9.3678(19) | 91.17(3) |
| 380 | $P2_1/n$ | 6.5382(4) | 7.4743(5) | 9.3541(5) | 91.25(5) |

The ferroelectricity of the SPS single crystal was confirmed by the temperature-dependent dielectric permittivity and hysteresis loop measurements, as shown in Figure 1. In Figure 1a and 1b, a prominent dielectric anomaly was observed at 339 K corresponding to Curie temperature $T_c$, which is consistent with previous work [31]. In the vicinity of $T_c$, the reciprocal dielectric permittivity $\varepsilon'$ follows the Curie-Weiss law of ferroelectric materials, $\varepsilon' = C/(T - T_0)$, as confirmed by the linear relationship between the reciprocal dielectric permittivity and the temperature, as shown in Figure 1c. The fitting parameters are Curie-Weiss constant $C=1.04\times10^4$ K and Curie-Weiss temperature $T_0$=325.8 K at 1 kHz. The ferroelectricity of SPS crystal was further confirmed by the typical polarization-electric field hysteresis at room temperature, as shown in Figure 1d. It is found that the polarization increases with a decreasing frequency of the electric field, but the coercive field remains almost unchanged. The sample on the (011) surface shows a spontaneous polarization ($P_s$) of ~4.8 μC/cm² at a frequency of 20 Hz.

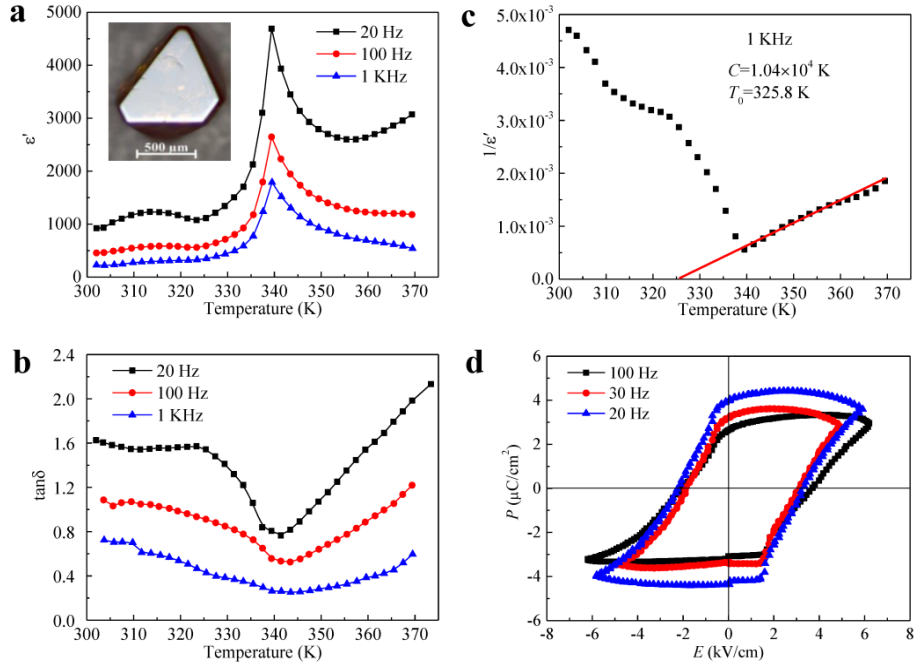

**Figure 1 Dielectric and ferroelectric properties of SPS single crystal.** The temperature dependencies of **(a)** dielectric permittivity $\varepsilon'$ and **(b)** dielectric loss tan$\delta$ for SPS crystal measured at 20 Hz, 100 Hz and 1 kHz. The inset of (a) is a photograph of SPS crystal. **(c)** The temperature dependencies of the reciprocal dielectric permittivity $1/\varepsilon'$ at 1 kHz. The data is linearly fitted following the Curie-Weiss law. **(d)** Frequency-dependent polarization versus electric field hysteresis loops of SPS crystal at 300 K.

We also used PFM to measure the local piezoelectric response and the ferroelectric domain structure. The out-of-plane (OP) phase and amplitude images in Figure 2a and 2b show two distinct regions with 180° phase difference, corresponding to domains with upward (small size domain) and downward (large size domain) polarization components perpendicular to the (011) surface. Furthermore, the in-plane (IP) phase image (Figure 2c) also exhibits two-color tones, corresponding to two opposite IP polarization components. The presence of IP polarization can be well explained by schematic illustration in Figure 2e that (011) surface is not perpendicular to the direction of spontaneous polarization in SPS single crystal, which will be discussed later. More details on various OP and IP PFM images on different locations can be found in Supporting Information Figure S2.

To make sure that the measured piezoresponse arises primarily from linear piezoelectricity instead of electrostatic effects, electrochemical dipoles, or ionic motions, which is known to cause apparent piezoresponse as well [32,33], we compare the first and second harmonic piezoresponse [34] versus the *ac* excitation (Figure 2f). The first harmonic response of SPS exhibits a clear linear behavior with increased sample voltage. SPS crystal has a significant first harmonic response that is much higher than the second harmonic one, suggesting that the electromechanical coupling in SPS is predominantly linear piezoelectric, though extrinsic nonlinear effects also exist. Moreover, the local phase and amplitude loops of SPS crystal was measured using switching spectroscopy PFM (SS-PFM), as shown in Figure 2g and 2h. The hysteresis loop of the sample shows butterfly characteristic and is symmetric. The effective $d_{33}$ value was estimated from the positive linear part of amplitude loop, giving a value about 28.2 ± 1.2 pm/V.

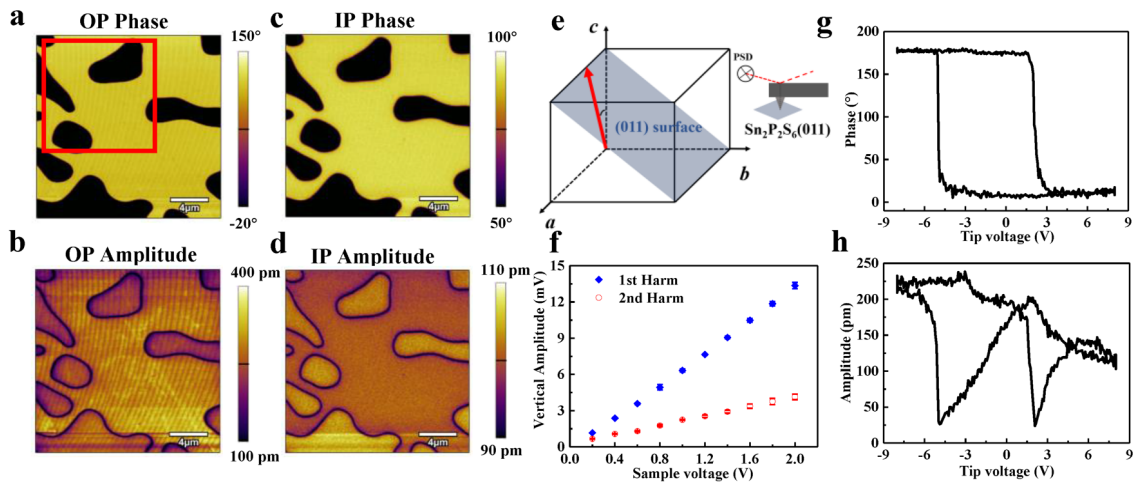

**Figure 2 Ferroelectricity of SPS single crystal by scanning probe microscope.** **(a)** and **(b)** show OP phase and amplitude, the corresponding IP phase and amplitude are displayed in **(c)** and **(d)**. **(e)** Schematic of polarization direction, scanning surface and experimental set-up. The (011) surface (gray colored) is scanned and polarization vector (red arrow) lies in the *ac*-plane and slightly deviates from the *c*-axis. **(f)** Comparison of first and second

harmonic piezoresponse versus sample voltage. **(g, h)** The corresponding PFM phase and amplitude hysteresis loops during the switching process for SPS crystal.

To further investigate the electrical conductivity, we performed a combination of c-AFM and sMIM technique study of both domains and domain walls on the (011) surface. Figure 3a shows the topography together with c-AFM images on an enlarged area in Figure 2a (labeled as red solid box). It is obvious that the current mapping is independent from the topographic feature. As the sample voltage is swapping from -10 to 6 V, the corresponding c-AFM images are obtained consecutively. As shown in Figure 3a, for positive sample voltages, we first detect domain wall current at voltages higher than 5 V, while no domain wall current is detected even at much higher negative sample voltages. Besides, there are two important features from the c-AFM images: (1) the domain walls are more conductive than the domains, and (2) two domains with opposite polarizations separated by the CDW exhibit distinguishable conductivity. Both features are evident from the line profile in Figure 3b and I-V curves in Figure 3c. The local I-V curves in Figure 3c are performed on different spots in different domains and domain walls (as shown in Figure 3b) with sample voltage from -10 to 10 V. We define the domains with upward polarization components perpendicular to the (011) surface as up domains, and the domains with downward polarization components as down domains. The domain wall conductivity can reach up to 1 nA with 10 V sample voltage, while conductivity of down domains up to 0.6 nA and up domains up to 0.3 nA, as shown in Figure 3c. It is also clear from Figure 3c that I-V curves are strongly nonlinear and asymmetric, and the positive current is much larger than the negative current, showing the obvious diode-like conducting behavior. In addition, the domain wall conductance can be higher than the domains conductance, varying by more than one order of magnitude. For positive sample voltages, we do not observe domain wall current until much higher voltages (5 V), which is consistent with the result of c-AFM measurement in Figure 3a.

It is also worth noting that the scanning rate has significant influence on the sample surface quality. All the c-AFM images are obtained with scanning rate 5 or 10 Hz, which preserves the surface quality. On the contrary, with low scanning rate 1 Hz and large applied *dc* voltage ($V_{dc}$>6 V), the surface is easily damaged due to the oxidization of the sample surface [35], as shown in Figure S3 in Supporting Information. We further employed sMIM [36,37] to confirm the conductivity of SPS crystal. The principle of sMIM is different from c-AFM, which directly uses high frequency (~GHz) to detect the local permittivity and conductivity, avoiding the damage of the sample surface. Figure 3d shows the simultaneously taken topography and sMIM imaginary part images of SPS sample. Domain walls and domains in single crystal with very smooth surfaces can be imaged directly, excluding the interference of topographic crosstalk to sMIM signals. Bright (conducting) areas appear on domain walls, which further confirms remarkable domain wall conductivity, as shown in Figure 3d. Note that there is no conductivity contrast between two different domains in sMIM image, which is different from the results of c-AFM measurement, and we will discuss this later.

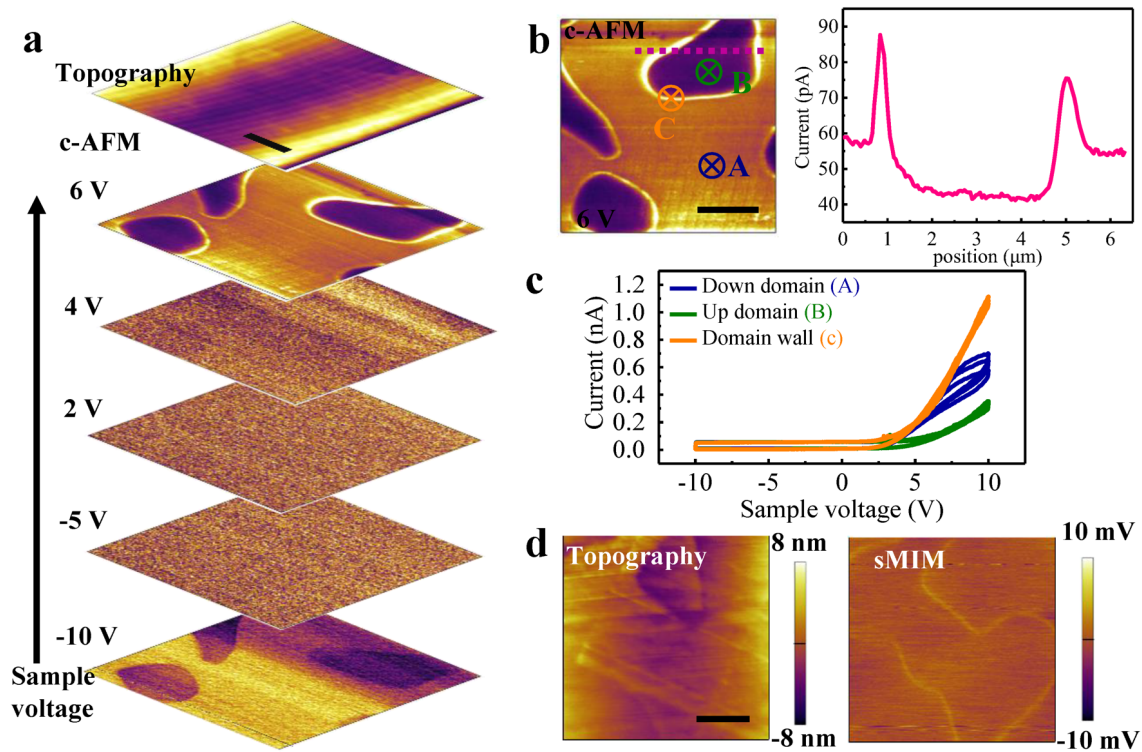

**Figure 3 Conduction at domain walls in SPS single crystal. (a)** Topography and c-AFM images of SPS (011) surface taken with different applied voltages. Brighter regions correspond to larger absolute current value. **(b)** The c-AFM image at 6 V with current profile along the pink dashed line. Three different spots on different domains and domain wall, as labeled by A, B and C, are corresponding to the local I-V curve measurements in (c). **(c)** The I-V curves on three different spots. **(d)** sMIM and corresponding topography images. The scale bar is 2 μm.

Guided by the experimental observations discussed above, now we will investigate the mechanism of the conductive domain walls in SPS crystal, as well as different conductive behaviors of two different domains from c-AFM and sMIM. Firstly, we calculated the polarization and analyzed domain configurations to get insight into the CDW. Based on the XRD refined structure of ferroelectric phase, as shown in Figure 4a and 4b, we found that the $(P_2S_6)^{4-}$ anionic units are locally centric symmetry in ferroelectric phase, while the Sn centred polyhedrons are locally acentric symmetry and the displacement of Sn relative to the center of S polyhedron is summarized in Table S2. From Table S2, it can be found that Sn has off-centering displacement in each individual polyhedron for paraelectric and ferroelectric phases. But for paraelectric phase, these off-centering displacements cancel out and therefore showing zero net polarization. For ferroelectric phase, only the displacements along *b* direction cancel out but Sn off-centering displacements are nonzero along *a* and *c* direction, showing net polarization in *ac*-plane. The off-centering displacements of Sn atoms are remarked as red arrows in Figure 4a and 4b. Moreover, we calculated the polarization by the Berry-phase method from density functional theory simulation. The net polarization is indeed in *ac*-plane, rotating 9.31° from *c*-axis with *a* component 2.29 μC/cm$^2$ and *c* component 16.24 μC/cm$^2$ (Figure 2e). With the inclined domain wall orientations in SPS[25-26], this net polarization perpendicular to (011) plane $P_\perp$ could be normal to the domain walls and form the tail-to-tail polarization distribution on the domain walls, as illustrated in Figure 4c. Besides, SPS crystal has semiconducting property with hole (*p*-type) electric conductivity at room temperature [38]. Therefore, this tail-to-tail polarization leads to the negative bound charges accumulated on domain walls and attracts the hole carriers (SPS is a *p*-type semiconductor), which results in the conducting behaviors on the domain walls.

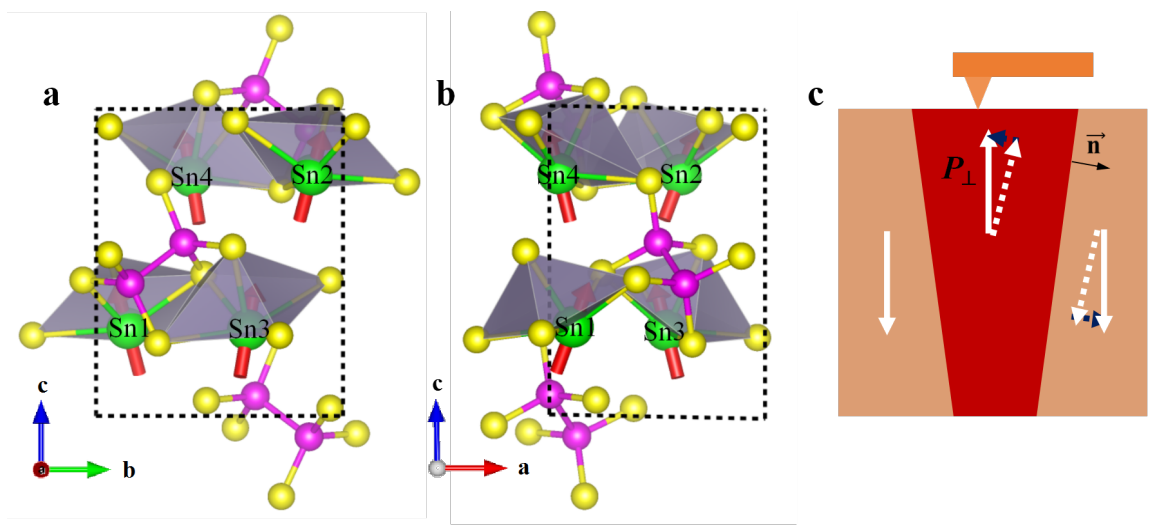

**Figure 4 Crystal structure and schematic of inclined domain walls of SPS single crystal.** Crystal structure of ferroelectric phase SPS (space group: $P2_1/n$) in the view along *a*-axis and *b*-axis are displayed in **(a)** and **(b)**,

respectively. The off-centering displacements of Sn atoms in ferroelectric SPS are remarked as red arrows. There are two inequivalent Sn atom groups: one group includes Sn1/Sn2 atoms, the other one includes Sn3/Sn4 atoms. Sn, P, and S atoms are in green, pink, and yellow, respectively. **(c)** Schematic illustration of inclined domain walls in SPS crystal, the net polarization perpendicular to (011) plane $P_\perp$ are labeled by white solid arrows, white and blue dashed arrows mean the parallel and perpendicular components respect to the domain wall, $n$ is domain wall normal vector.

Secondly, the asymmetry I-V behavior and distinguishable electrical conductivity of domains separated by CDW in c-AFM image result from a Schottky-like barrier at the tip-sample interface which is modified by the local ferroelectric polarization [39-41]. Figure 5 shows the schematic diagrams of interfacial band, where $\varphi_M$ and $\varphi_S$ represent the work function of tip and SPS crystal, respectively, and $\varphi_b$ represents the Schottky barrier height for conduction. As shown in Figure 5a, the interfacial polarization charge with opposite signs is produced on the up and down domains, which bends the conduction band ($E_C$) and valence band ($E_V$) in an opposite way and causes the difference of Schottky barrier $\Delta\varphi_b$ at the interface.

Previous study confirmed the hole-type conductivity at room temperature in SPS crystal [38]. Therefore, when conductive tip is in contact with SPS sample with ferroelectric domains, a Schottky barrier for holes is formed at interface to maintain a uniform Fermi energy $E_F$, as shown in Figure 5b. Under positive sample voltage (negative tip voltage), as shown in Figure 5c, the Schottky barrier for hole transport from SPS sample to tip is relatively low under positive sample voltage (negative tip voltage) and the barrier further decreases with increasing voltage. While Schottky barrier for hole transport from tip to SPS crystal is very high under negative sample voltage (positive tip voltage) and the barrier further increases with increasing voltage (Figure 5d). Therefore, the conduction under positive sample voltage is much larger than under negative sample voltage and I-V curves are asymmetry, as shown in Figure 3c. A lower φ$_b$ is produced at tip-down domain interface than tip-up domain interface and well accounts for higher conductivity for down domains than up domain using c-AFM (*dc* measurement). On the contrary, sMIM measurement is conducted at *ac* high frequencies (~GHz) and those charge carriers blocked by the Schottky barriers at *dc* measurement are facilitated to contribute to the conduction by oscillating between the barriers [42]. Therefore, in sMIM measurement, conduction is insensitive to the Schottky barrier at contact interface and no conductivity contrast between two different domains is observed.

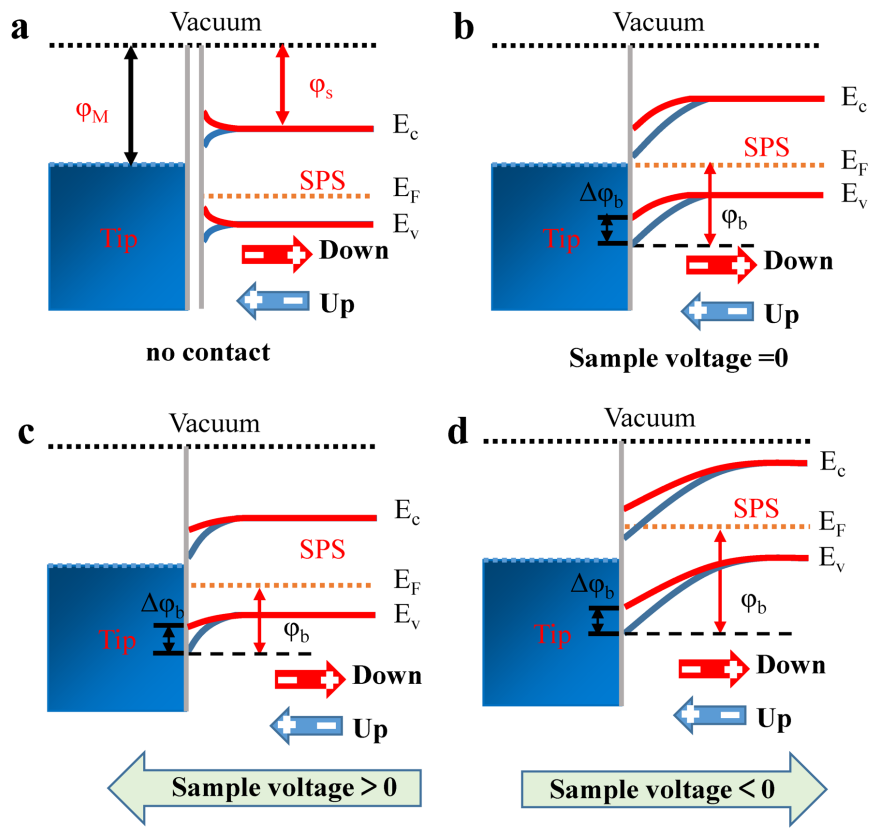

**Figure 5 Schematic diagrams of interfacial band between conductive Ti/Ir tip and ferroelectric domains. (a)** No contact between tip and SPS sample. The presence of polarization bends the conduction (valence) band. **(b)** Sample voltage = 0 (zero voltage); **(c)** Sample voltage > 0 (positive voltage); **(d)** Sample voltage < 0 (negative voltage). φ$_M$ (φ$_S$) is the work function of tip (SPS sample). E$_c$, E$_v$ and E$_F$ are energy level of conduction band, valence band and Fermi energy of SPS crystal, respectively. Blue and red thick arrows represent the upward and downward polarization components perpendicular to the (011) surface, respectively.

## 3. Conclusion

In summary, we demonstrated that the existence of conductivity domain walls in the non-oxide ferroelectric $Sn_2P_2S_6$ single crystal by using a combination of piezoresponse force microscope, conductive atomic force microscope and scanning microwave impedance microscope. Moreover, the two different domains with opposite polarization directions exhibit distinguishable electrical conduction, which is caused by the different band bending behavior due to interfacial charge with opposite signs. Though the mechanism for the domain wall conductivity is similar as the oxide ferroelectric possesses, such room temperature conductive domain walls in non-oxide ferroelectric provides a brand-new platform for the investigation of oxygen-related domain walls properties, including the oxygen defects induced conductivity variations. Our results pave the way for understanding conductivity of the domains and domain walls in non-oxide ferroelectrics and offer opportunities for a new kind of nanoscale conduction channel in multifunctional devices.

**Experimental Section**

**Sample preparation and characterization**

Single crystals of SPS were synthesized by chemical vapor transport (CVT) in the two-zone furnace. Stoichiometric amounts of Sn, P, and S were ground together and then transferred to quartz tube. The powders were evacuated, sealed, and placed in a two-zone furnace (650-600 °C). The pressure inside the tube was pumped down to $3 \times 10^{-4}$ Pa. After one week of heating, the SPS sample was cooled down to room temperature in the growth zone. The crystals formed as yellowish-brown polyhedra, with irregular shapes and up to about $5 \times 4 \times 4$ mm$^3$. From the batch, small crystals (orange/yellow, transparent) suitable for single crystal X-ray diffraction experiments were selected. To study the crystal structure of SPS sample, single-crystal X-ray diffraction data were collected on a Rigaku Oxford XtaLAB PRO diffractometer with graphite-monochromated Mo $K_\alpha$ radiation ($\lambda$=0.71073 Å) at temperatures above and below the ferroelectric transition temperature. The crystal surface of SPS crystal was identified by X-ray diffraction (XRD) collected from a Bruker D8 Advance x-ray diffractometer using Cu $K_\alpha$ radiation at room temperature. Scanning electron microscopy (SEM) photographs were carried out on FEI Nano 450 microscope at a voltage of 20 kV. The atomic composition of SPS single crystal was checked by energy dispersive X-ray spectroscopy (EDS, Oxford X-Max 50).

**Dielectric and ferroelectric properties measurement**

The dielectric property was measured by a precision impedance analyzer (WK, 6500B, UK) at the temperature range from 300 to 370 K. The dielectric permittivity as a function of frequency was measured with an *ac* excitation voltage of 0.5 V. Ferroelectric hysteresis loops at room temperature were measured using a standard ferroelectric tester (Radiant Technology, Precision Premier II, USA) as a function of frequency from 0.01 - 20 kHz.

**Scanning probe microscope measurement**

PFM measurement was performed using a commercial atomic force microscope (Asylum Research MFP-3D) with Ti/Ir-coated Si cantilever tips and diamond-coated Si cantilever tips, respectively. In resonance-enhanced mode, a soft tip with a spring constant of ~2.8 N m$^{-1}$ was driven with an *ac* voltage ($V_{ac}$ = 0.5-1 V) under the tip-sample contact resonant frequency (~320 kHz). The OP and IP PFM were acquired at the drive frequency of ~320 kHz and ~730 kHz in Vector PFM mode, respectively. SS-PFM was performed by measuring hysteresis loops in the OP piezoelectric signals using DART mode. c-AFM was performed in contact mode with the conductive AFM tip as the top electrode. SPS is also photo-sensitive material, therefore the measurement environment is without light to avoid the additional contributions. sMIM measurement in this work is based on an AFM platform. The customized shielded cantilevers are commercially available from PrimeNano Inc.

**Density Functional Theory (DFT) Calculation**

All calculations in this work were carried out using the Vienna Ab initio Simulation Package (VASP) based on DFT[43]. For the exchange-correlation functional, PBEsol[44] functional of generalized gradient approximation was adopted. In the calculation, kinetic energy cutoff of 500 eV for the plane wave expansion and $6 \times 6 \times 4$ Monkhorst-Pack k-point grids were were set for the bulk SPS. The energy and force convergence criterions are $10^{-6}$ eV and 0.5 meV/Å during the relaxation, respectively. Additionally, the Berry-phase method was employed to calculate the polarization for bulk SPS[45-47].

**Acknowledgements**

The work at Beijing Institute of Technology is supported by National Natural Science Foundation of China with Grant Nos. 11572040, 11604011, 11804023, the Beijing Natural Science Foundation (Z190011) and the China Postdoctoral Science Foundation with Grant No. 2018M641205. X. W. also acknowledges the National Key Research and Development Program of China (2019YFA0307900) and Beijing Institute of Technology Research Fund Program for Young Scholars. The authors are grateful for the fruitful discussions with Jing Wang from Beijing Institute of Technology.

**Author contributions**

X.W. and J.H. designed and supervised the experiments. J.D., X.J., M.W. and X.W. prepared the samples, performed PFM and c-AFM experiments. J.D. and W.Z. carried out dielectric property characterization. J.D. and Z.G. performed the sMIM characterization. Y.L. and P.L. performed DFT calculations. XRD experiments are done by Y.L., S.X., J.W., Z.Y. and T.X. All authors discussed the results and commented on the manuscript.

**Notes**

The authors declare no competing financial interest.

# References

1. G. Catalan, J. Seidel, R. Ramesh, J. F. Scott, *Rev. Mod. Phys.,* 2012, **84**, 119-156.

2. J. A. Mundy, J. Schaab, Y. Kumagai, A. Cano, M. Stengel, I. P. Krug, D. M. Gottlob, H. Dog Anay, M. E. Holtz, R. Held, Z. Yan, E. Bourret, C. M. Schneider, D. G. Schlom, D. A. Muller, R. Ramesh, N. A. Spaldin, D. Meier, *Nat. Mater.,* 2017, **16**, 622-627.

3. J. Schaab, S. H. Skjaervo, S. Krohns, X. Dai, M. E. Holtz, A. Cano, M. Lilienblum, Z. Yan, E. Bourret, D. A. Muller, M. Fiebig, S. M. Selbach, D. Meier, *Nat. Nanotechnol.,* 2018, **13**, 1028-1034.

4. P. S. Bednyakov, B. I. Sturman, T. Sluka, A. K. Tagantsev, P. V. Yudin, *npj Computational Mater.,* 2018, **4**, 1-11.

5. S. Farokhipoor, B. Noheda, *Phys. Rev. Lett.,* 2011, **107**, 127601.

6. T. Sluka, A. K. Tagantsev, P. Bednyakov, N. Setter, *Nat. Commun.,* 2013, **4**, 1808.

7. D. Meier, J. Seidel, A. Cano, K. Delaney, Y. Kumagai, M. Mostovoy, N. A. Spaldin, R. Ramesh, M. Fiebig, *Nat. Mater.,* 2012, **11**, 284-288.

8. J. Seidel, L. W. Martin, Q. He, Q. Zhan, Y. H. Chu, A. Rother, M. E. Hawkridge, P. Maksymovych, P. Yu, M. Gajek, N. Balke, S. V. Kalinin, S. Gemming, F. Wang, G. Catalan, J. F. Scott, N. A. Spaldin, J. Orenstein, R. Ramesh, *Nat. Mater.,* 2009, **8**, 229-234.

9. M. Schroder, A. Haussmann, A. Thiessen, E. Soergel, T. Woike, L. M. Eng, *Adv. Funct. Mater.* 2012, **22**, 3936-3944.

10. W. Wu, Y. Horibe, N. Lee, S. W. Cheong, J. R. Guest, *Phys. Rev. Lett.,* 2012, **108**, 077203.

11. Y. S. Oh, X. Luo, F. T. Huang, Y. Z. Wang, S. W. Cheong, *Nat. Mater.,* 2015, **14**, 407-413.

12. M. Schrade, N. Maso, A. Perejon, L. A. Perez-Maqueda, A. R. West, *J. Mater. Chem. C,* 2017, **5**, 10077-10086.

13. X. Wang, D. Yang, H.-M. Zhang, C. Song, J. Wang, G. Tan, R. Zheng, S. Dong, S.-W. Cheong, J. Zhang, *Phys. Rev. B,* 2019, **99**, 054106.

14. Y. Du, X. L. Wang, D. P. Chen, S. X. Dou, Z. X. Cheng, M. Higgins, G. Wallace, J. Y. Wang, *Appl. Phys. Lett.,* 2011, **99**, 252107.

15. J. Seidel, P. Maksymovych, Y. Batra, A. Katan, S. Y. Yang, Q. He, A. P. Baddorf, S. V. Kalinin, C. H. Yang, J. C. Yang, Y. H. Chu, E. K. H Salje, H. Wormeester, M. Salmeron, R. Ramesh, *Phys. Rev. Lett.,* 2010, **105**, 197603.

16. A. L. Solomon, THIOUREA, A NEW FERROELECTRIC. *Phys. Rev.,* 1956, **104**, 1191-1191.

17. G. J. Goldsmith, J. G. White, *J. Chem. Phys.,* 1959, **31**, 1175-1187.

18. S. Sawada, Y. Shiroishi, A. Yamamoto, M. Takashige, M. Matsuo, *J. Phys. Soc. Jpn.,* 1977, **43**, 2099-2100.

19. M. A. Susner, M. Chyasnavichyus, M. A. McGuire, P. Ganesh, P. Maksymovych, *Adv. Mater.,* 2017, **29**, 1062852.

20. F. Liu, L. You, K. L. Seyler, X. Li, P. Yu, J. Lin, X. Wang, J. Zhou, H. Wang, H. He, S. T. Pantelides, W. Zhou, P. Sharma, X. Xu, P. M. Ajayan, J. Wang, Z. Liu, *Nat. Commun.,* 2016, **7**, 12357.

21. A. Belianinov, Q. He, A. Dziaugys, P. Maksymovych, E. Eliseev, A. Borisevich, A. Morozovska, J. Banys, Y. Vysochanskii, S. V. Kalinin, *Nano Lett.,* 2015, **15**, 3808-14.

22. M. Chyasnavichyus, M. A Susner, A. V. Ievlev, E. A. Eliseev, S. V. Kalinin, N. Balke, A. N. Morozovska, M. A. McGuire, P. Maksymovych, *Appl. Phys. Lett.,* 2016, **109**, 172901.

23. L. You, Y. Zhang, S. Zhou, A. Chaturvedi, S. A. Morris, F. Liu, L. Chang, D. Ichinose, H. Funakubo, W. Hu, *Sci. Adv.,* 2019, **5**, eaav3780.

24. J. M. Deng, Y. Y. Liu, M. Q. Li, S. Xu, Y. Z. Lun, P. Lv, T. L. Xia, P. Gao, X. Y. Wang, J. W. Hong, *Small,* 2020, **16**, 1904529.

25. D. I. Kaynts, A. A. Grabar, M. I. Gurzan, A. A. Horvat, *Ferroelectrics,* 2004, **304**, 187-191.

26. A. A. Grabar, I. V. Kedyk, I. M. Stoiak, Yu. M. Vysochanskii, *Ferroelectrics,* 2001, **254**, 285-293.

27. C. D. Carpentier, R. Nitsche, *Mater. Res. Bull.,* 1974, **9**, 1097-1100.

28. I. Zamaraite, R. Yevych, A. Dziaugys, A. Molnar, J. Banys, S. Svirskas, Y. Vysochanskii, *Phys. Rev. Appl.,* 2018, **10**, 034017.

29. R. M. Yevych, Y. M. Vysochanskii, *Ferroelectrics,* 2011, **412**, 38-44.

30. Y. C. Rong, K. Lin, F. M. Guo, R. H. Kou, J. Chen, Y. Ren, X. R. Xing, *J. Phys. Chem. C,* 2017, **121**, 1832-1837.

31. J. Grigas, V. Kalesinskas, S. Lapinskas, W. Paprotny, *Ferroelectrics,* 1988, **80**, 225-228.

32. S. V. Kalinin, E. Karapetian, M. Kachanov, *Phys. Rev. B,* 2004, **70**, 184101.

33. Q. N. Chen, Y. Ou, F. Y. Ma, J. Y. Li, *Appl. Phys. Lett.,* 2014, **104**, 242907.

34. P. Jiang, F. Yan, E. N. Esfahani, S. H. Xie, D. F. Zou, X. Y. Liu, H. R. Zheng, J. Y. Li, *Acs Biomater. Sci. & Engineer.,* 2017, **3**, 1827-1835.

35. P. Avouris, T. Hertel, R. Martel, *Appl. Phys. Lett.,* 1997, **71**, 285-287.

36. K. Lai, W. Kundhikanjana, M. Kelly, Z. X. Shen, *Rev. Sci. Instrum.,* 2008, **79**, 063703.

37. K. Lai, W. Kundhikanjana, M .A. Kelly, Z. X. Shen, *Appl. Phys. Lett.,* 2008, **93**, 123105.

38. Yu. Vysochanskii, K. Glukhov, M. Maior, K. Fedyo, A. Kohutych, V. Betsa, I. Prits, M. Gurzan, *Ferroelectrics,* 2011, **418**, 124-133.

39. L. Pintilie, M. Alexe, *J. Appl. Phys.,* 2005, **98**, 124103.

40. L. Pintilie, I. Boerasu, M. J. M. Gomes, T. Zhao, R. Ramesh, M. Alexe, *J. Appl. Phys.,* 2005, **98**, 124104.

41. W. D. Wu, J. R. Guest, Y. Horibe, S. Park, T. Choi, S. W. Cheong, M. Bode, *Phys. Rev. Lett.,* 2010, **104**, 217601.

42. A. Tselev, P. Yu, Y. Cao, L. Dedon, L. Martin, S. Kalinin, P. Maksymovych, *Nat. Commun.,* 2016, **7**, 1-9.

43. G. Kresse, D. Joubert, *Phys. Rev. B,* 1999, **59**, 1758.

44. J. P. Perdew, A. Ruzsinszky, G. I. Csonka, O. A. Vydrov, G. E. Scuseria, L. A. Constantin, X. Zhou, K. Burke, *Phys. Rev. Lett.,* 2008, **100**, 136406.

45. X. Wu, D. Vanderbilt, D. R. Hamann, *Phys.Rev. B,* 2005, **72**, 035105.

46. R. D. King-Smith, D. Vanderbilt, *Phys. Rev. B,* 1993, **47**, 1651.

47. D. Vanderbilt, *J. Phys. Chem. Solids,* 2000, **61**, 147-151.

# Supporting Information

## Conductive Domain Walls in Non-Oxide Ferroelectrics $Sn_2P_2S_6$

Jianming Deng[1,#], Xingan Jiang[1,#], Yanyu Liu[1], Wei Zhao[1], Yun Li[2], Ziyan Gao[1], Peng Lv[1], Sheng Xu[3], Tian-Long Xia[3], Jinchen Wang[3], Meixia Wu[4], Zishuo Yao[2], Xueyun Wang[1*], Jiawang Hong[1*]

[1]School of Aerospace Engineering, Beijing Institute of Technology, Beijing, 100081, China
[2]School of Chemistry and Chemical Engineering, Beijing Institute of Technology, Beijing, 100081, China
[3]Department of Physics and Beijing Key Laboratory of Opto-electronic Functional Materials & Micro-nano Devices, Renmin University of China, Beijing 100871, China
[4]Key Laboratory of Bioinorganic and Synthetic Chemistry of Ministry of Education, School of Chemistry, Sun Yat-Sen University, Guangzhou 510275, P. R. China.

[#]These authors contribute equally to this work.
[*]Corresponding e-mails: xueyun@bit.edu.cn, hongjw@bit.edu.cn

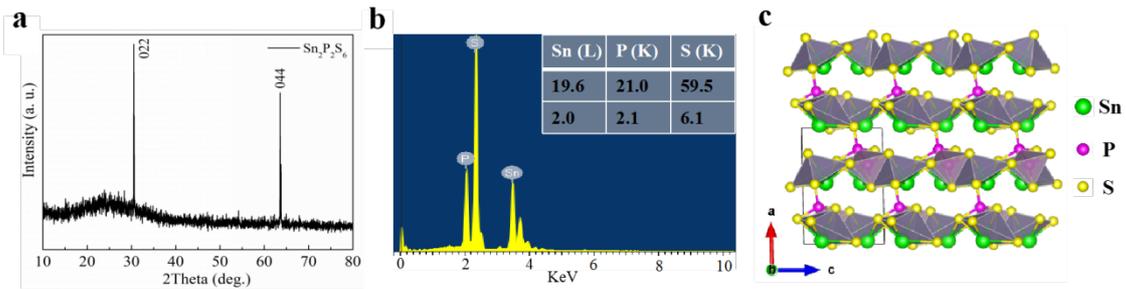

Figure S1. Crystal structure and Characterization of SPS single crystal. (a) The X-ray diffraction for SPS. (b) EDS spectrum of the SPS sample. (c) Side view of the crystal structure of ferroelectric SPS (space group: *Pn*). The EDS result for SPS indicates our sample with a high quality.

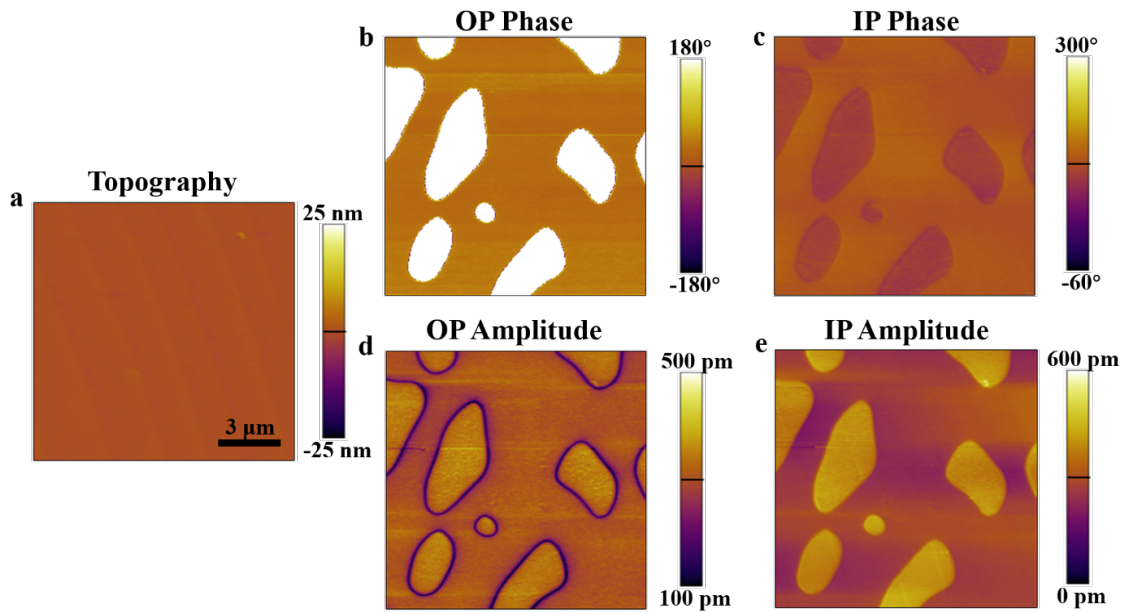

Figure S2. Piezoresponse force microscopy of SPS sample. (a) The topography, (b) OP phase, (d) OP amplitude and the corresponding (c) IP phase, (e) IP amplitude images of SPS crystal. The scale bar is 3 μm. The OP phase image shows two-color tones with a contrast of 180°, corresponding to domains with up and down polarization components perpendicular to the (011) surface, whereas the domain walls appear as darker lines in the OP amplitude image (Figure S2d). Interestingly, phase contrast is observed for SPS in IP direction, as shown in Figure S2c.

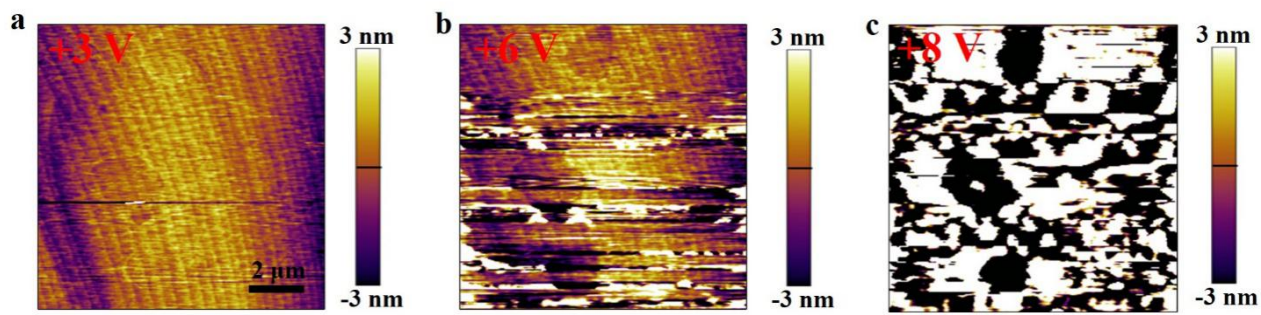

Figure S3. Topography with different voltage at 1 Hz scanning rate. The results indicate that slow scan with high voltage (more than 6 V) severely damage the surface of SPS sample.

Table S1 Rietveld refinement results of the X-ray diffraction data for SPS at different temperature.

| | 163 K | 300 K | 380 K |
|---|---|---|---|
| Space group | $Pn$ | $Pn$ | $P2_1/n$ |
| Atomic position | x, y, z | x, y, z | x, y, z |
| Sn1 | 0.1093(10), 0.1271(10), 0.2852(7) | 0.1093(8), 0.1254(9), 0.2793(7) | 0.958945 0.369282 0.742917 |
| Sn2 | 0.5423(11), 0.6141(9), 0.2807(8) | 0.5373(9), 0.6149(9), 0.2729(7) | |
| P1 | 0.5046(4), 0.3553(4), 0.5652(3) | 0.5058(2), 0.3559(3), 0.5660(2) | 0.439331 0.391408 0.567109 |
| P2 | 0.6270(5), 0.1377(3), 0.4319(3) | 0.6264(3), 0.1388(3), 0.4319(2) | |
| S1 | 0.6633(4), 0.2496(3), 0.2378(3) | 0.6650(3), 0.2490(3), 0.2372(2) | 0.177500 0.308817 0.467208 |
| S2 | 0.2436(4), 0.4361(4), 0.4615(3) | 0.2443(3), 0.4370(3), 0.4625(2) | |
| S3 | 0.7162(4), 0.5530(3), 0.5501(3) | 0.7176(3), 0.5522(3), 0.5527(2) | 0.655675 0.197597 0.556879 |
| S4 | 0.4641(4), 0.2569(3), 0.7626(3) | 0.4651(3), 0.2556(3), 0.7630(2) | |
| S5 | 0.8906(4), 0.0525(4), 0.5255(3) | 0.8904(3), 0.0546(3), 0.5270(2) | 0.398877 0.497664 0.763006 |
| S6 | 0.4047(4), -0.0533(3), 0.4426(3) | 0.4071(2), -0.0525(3), 0.4423(2) | |
| R factor [a] | $R_{wp}$=0.099, $R_p$=0.039 | $R_{wp}$=0.066, $R_p$=0.027 | $R_{wp}$=0.1187 $R_p$=0.0475 |

[a] $R_p$ is sum($|I_0-I_C|$)/sum($I_0$), and $R_{wp}$ is weighted R factors for X-ray diffraction data

Table S2 Off-centering displacements of Sn atom relative to the center of the coordinated S atoms in polyhedron.

| Displacement | 163 K | | | 300 K | | | 380 K | | |
|---|---|---|---|---|---|---|---|---|---|
| | a | b | OP [a] | a | b | OP [a] | a | b | OP [a] |
| Sn1 (Å) | 0.495 | 0.310 | -0.945 | 0.489 | 0.291 | -1.002 | 0.282 | 0.976 | -0.418 |
| Sn2 (Å) | 0.495 | -0.310 | -0.945 | 0.489 | -0.291 | -1.002 | 0.282 | -0.976 | -0.418 |
| Sn3 (Å) | -0.365 | 0.212 | -1.033 | -0.407 | 0.216 | -1.110 | -0.282 | 0.976 | 0.418 |
| Sn4 (Å) | -0.365 | -0.212 | -1.033 | -0.407 | -0.216 | -1.110 | -0.282 | -0.976 | 0.418 |

[a]OP represents the direction of perpendicular to *ab*-plane